\documentclass[usenatbib]{mn2e}
\usepackage[]{graphicx}
\usepackage{amsmath}

\title[The OGLE-III planet detection efficiency]{The OGLE-III planet detection efficiency from six years of microlensing observations (2003-2008)}
\author[Y. Tsapras]
{\parbox[t]{\textwidth}{ 
Y. Tsapras$^{1,2}$, M. Hundertmark$^{3}$, {\L}. Wyrzykowski$^{4}$, K. Horne$^{7}$, A. Udalski$^{4}$\\
 {\it and}\\
C. Snodgrass$^{6}$, R. Street$^{2}$, D. M. Bramich$^{5}$, M. Dominik$^{7}$, V. Bozza$^{11,12}$, R. Figuera Jaimes$^{7,14}$, N. Kains$^{13}$,
J. Skowron$^{4}$,  M. K. Szyma{\'n}ski$^{4}$, 
G. Pietrzy{\'n}ski$^{4}$, I. Soszy{\'n}ski$^{4}$, K. Ulaczyk$^{4,9}$,  S. Koz{\l}owski$^{4}$,  
P. Pietrukowicz$^{4}$, R. Poleski$^{4,8}$\\
}
\\
$^{1}$Astronomisches Rechen-Institut, Zentrum f{\"u}r Astronomie der Universit{\"a}t Heidelberg (ZAH), 69120 Heidelberg, Germany\\
$^{2}$Las Cumbres Observatory Global Telescope Network, 6740 Cortona Drive, suite 102, Goleta, CA 93117, USA\\
$^{3}$Niels Bohr Institutet,K{\o}bnhavns Universitet,Juliane Maries Vej 30, 2100 K{\o}benhavn {\O}, Denmark\\
$^{4}$Warsaw University Astronomical Observatory, Al.~Ujazdowskie~4, 00-478~Warszawa, Poland \\
$^{5}$Qatar Environment and Energy Research Institute (QEERI), HBKU, Qatar Foundation, Doha, Qatar\\
$^{6}$Planetary and Space Sciences, Department of Physical Sciences, The Open University, Milton Keynes, MK7 6AA, UK\\
$^{7}$SUPA, School of Physics \& Astronomy, University of St Andrews, North Haugh, St Andrews KY16 9SS, UK\\
$^{8}$Department of Astronomy, Ohio State University, 140 W. 18th Ave., Columbus, OH 43210, USA\\
$^{9}$Department of Physics, University of Warwick, Coventry CV4 7AL, UK\\
$^{10}$Universidad de Concepci\'{o}n, Departamento de Fisica, Astronomy Group, Casilla 160-C, Concepci\'{o}n, Chile\\
$^{11}$Dipartimento di Fisica "E. R. Caianiello", Universit{`a} di Salerno, Via Giovanni Paolo II 132, Fisciano 84084, Italy\\
$^{12}$Istituto Nazionale di Fisica Nucleare, Sezione di Napoli, Italy\\
$^{13}$Space Telescope Institute, 3700 San Martin Drive, Baltimore, MD 21218, USA\\
$^{14}$European Southern Observatory, Karl-Schwarzschild-Straße 2, 85748 Garching bei M{\"u}nchen, Germany\\
}
\begin{document}

\date{Oct 2015}

\pagerange{\pageref{firstpage}--\pageref{lastpage}} \pubyear{2015}

\maketitle

\label{firstpage}

\begin{abstract}
We use six years (2003-2008) of OGLE-III microlensing observations to derive the survey detection efficiency for a range of planetary masses and projected distances from the host star. We perform an independent analysis of the microlensing light curves to extract the event parameters and compute the planet detection probability given the data. 2433 light curves satisfy our quality selection criteria and are retained for further processing. The aggregate of the  detection probabilities over the range explored yields the expected number of microlensing planet detections. We employ a Galactic model to convert this distribution from dimensionless to physical units, $\alpha/$AU and $m_{\oplus}$. The survey sensitivity to small planets is highest in the range 1-4AU, shifting to slightly larger separations for more massive ones.
\end{abstract}


\section{Introduction}
The prolific discoveries of planets orbiting distant stars over the past two decades
have radically changed the way we understand planetary systems. Current planet formation models involve protoplanets forming in a material-rich accretion disk surrounding the host star. These protoplanets co-evolve with the disk and may undergo orbital decay due to torque asymmetries in the surrounding disk material \citep{b1,b2}. 

The majority of exoplanet discoveries have been announced by radial velocity and transit surveys\footnote{http://exoplanet.eu} while thousands of new candidates were discovered by the Kepler mission \citep{b45}. These surveys are most sensitive to systems with massive planets in short orbits ({\it Hot Jupiters}) and require long survey lifetimes to detect signals from longer period planets. {However, significant progress has been made in recent years in discovering planets of a few $M_{\oplus}$ out to distnaces of $\sim$1 AU from their host stars \citep{b80,b81}}. Other search methods, such as direct imaging and microlensing, are also finding planets in a complementary region of parameter space that is largely unexplored by transits and radial velocity. The sensitivity of microlensing extends to small colder planets which orbit their stars at distances of a few astronomical units ($\sim$1-10 AU). The population of stars that the method explores are low-mass, typically M-dwarf, stars between the Solar system and the centre of the Galaxy. It can therefore be used as a tool to build a census of colder Galactic exoplanets. 

The region of microlensing sensitivity corresponds to a cold zone in protoplanetary disks that is more conducive to planet formation and which conveniently overlaps with the cold outer edge of the Habitable Zone. Current theories predict that small planetary bodies, made up of rock and ice, should be quite common in that region \citep{b4,b5}. As the expected orbital semi-major axis vs mass distribution of these planets depends on theoretical models of planetary formation and migration, microlensing discoveries play an important part in testing and refining these models. Ultimately, to gain an understanding of how planetary systems form and evolve, we need a sufficiently large sample of thousands of planet discoveries spanning the full range of parameter space from the very large to the very small and from the very close-in to more distant ones for a range of host star masses and metallicities. In this paper, we use the OGLE-III microlensing data from 2003 to 2008 to derive an estimate of the survey sensitivity to planetary companions to the lens star. Our method follows that of \citet{b6} and \citet{b7}. Throughout the paper we use the full notation for the OGLE events (e.g.OGLE-2004-BLG-490) but keep the abbreviated notation in figures and tables (e.g. OB04490) for convenience.

In Section 2 we discuss the microlensing method and provide a description of our fits to the OGLE light curves. The exploration of binary parameter space and derivation of detection probabilities are presented in Section 3. We conclude with a summary of this work in Section \ref{sec_sum}.

\section{The microlensing method}
\subsection{Microlensing by stars hosting planets}
Planetary microlensing was first mentioned as a possibility by \citet{b3}. Gravitational microlensing occurs when a foreground star ({\it lens}) happens to pass very close to our line of sight to a more distant star ({\it source}). The foreground star acts as a gravitational lens, bending the light coming from the more distant star, generating multiple distorted images of the source around the lensing star. The number of images depends on the number of lensing masses involved. A single lens produces two images, a binary lens three or five, depending on the location of the source relative to the lens. If the lens and source are perfectly aligned, the images merge and form a bright ring around the lens which is commonly referred to as the {\it Einstein ring}.

The radius of the Einstein ring is given by
\begin{equation}
R_E = \sqrt{\frac{4 G M D}{c^2}},
\end{equation}
where $M$ is the mass of the lens, $c$ the speed of light, $G$ the gravitational constant
and $ D = (D_{LS}D_L) /  D_S $, where $D_{LS}$ is the distance between the lens and the source, $D_S$ is the distance from the observer to the source and $D_L$ the distance from the observer to the lens.

In microlensing, the distances between the images generated by the lensing effect are too small to be resolved individually with current technology. What is actually observed during these events is an increase in the brightness of the source star as the lens moves closer to it, followed by a gradual dimming back to its normal brightness as the lens moves away. If the lensing star happens to host a planet, it may also act as a lens and further perturb the light coming from the source star. This results in short-lived anomalous features on the event light curve that reveal the presence of the planet. These anomalies typically last for a few days in the case of a Jupiter-mass planet down to a few hours for an Earth-mass planet.

Microlensing of stars by stars is a very rare phenomenon with only one in a million stars in the Galaxy being microlensed at any one time. However, two dedicated survey teams, OGLE\footnote{http://ogle.astrouw.edu.pl/} and MOA\footnote{http://www.phys.canterbury.ac.nz/moa/}, using 1m-class telescopes equipped with wide-field cameras, announce over 2000 on-line alerts of ongoing microlensing events every year \citep{b8,b9}.

A small subset of these events are selected for monitoring by follow-up teams (RoboNet\footnote{http://robonet.lcogt.net/}, PLANET\footnote{http://planet.iap.fr/}, MiNDSTeP\footnote{http://www.mindstep-science.org/}, $\mu$FUN\footnote{http://www.astronomy.ohio-state.edu/~microfun/}) to look for planetary deviations. Anomalies are generally recognized in real time and secondary alerts are issued \citep{b10} to trigger higher cadence observations that can confirm or disprove the planetary nature of the event. All teams pool their resources and observe the anomalous features from multiple telescopes in order to fully characterize the potential planet.

Since deviations produced by small Earth-mass planets only last for a few hours, it is crucial to respond promptly to these alerts and have many telescopes observe them from different longitudes. Overlapping observations from different sites are desirable as they facilitate easier inter-calibration between the datasets and independently confirm the anomalous nature of the signal.

\subsection{Microlensing planet detections}
Although there have been dozens of candidate planetary events detected by microlensing \citep{b11}, to date, only thirty-five microlensing planet discoveries have been published\footnote{www.exoplanet.eu, exoplanetarchive.ipac.caltech.edu},
two of which are multiple-planet systems. Characterization of these microlensing events entailed an extensive exploration of the parameter space where the viability of alternative models was assessed and where the planetary interpretation emerged as the only viable solution. Of these planets, some have masses between Jupiter and Saturn, a few have masses comparable to that of Neptune and three have masses that lie between 1.5 and 6 Earth masses.

For microlensing follow-up observing campaigns there are two main channels to planet discovery: a) concentrate on the rare high-magnification events exclusively where the probability of detecting giant planets approaches 100\% \citep{b22}, or b) maximize the chances of small-planet detection by adopting an observing plan that distributes the observations over a small number of high-interest events \citep{b24}.

Because in the high-magnification regime the planetary signature, associated with the central caustic, scales with the planet/star mass ratio $q$, high-magnification events are less sensitive to small planets; hence detections are biased towards more massive, Neptune and Jupiter-like planets. In the lower magnification regime, where the planetary caustic is responsible for the perturbations, the planet signatures scale more weakly, as $\sqrt{q}$, so high-cadence sampling on the wings of the light curve favours detection of smaller planets, down to just below the mass of the Earth.

\subsection{Estimating the planet abundance}
In order to draw conclusions about planetary populations, it is not enough to just detect planets, it is also necessary to understand the selection bias of the surveys. This calls for the adoption of a fully deterministic observing strategy. This requirement can be satisfied either by a combination of survey and follow-up observations that make use of a fully robotic system controlled by deterministic algorithms which prioritize and select the target events automatically, such as the approach being developed by RoboNet \citep{b23, b24}, or by performing sequential observations of Galactic bulge fields, monitoring millions of stars in survey mode with a cadence of ~15-20 minutes using wide-field cameras from multiple sites, which is the approach followed by the newly commissioned KMTNet project \cite{b65}. However, it is possible to start addressing this question using the existing pure survey data by OGLE or MOA, even though there are longitudinal gaps in the sampling since single sites are used for the observations.
\begin{table*}
\centering
 \begin{minipage}{126mm}
  \caption{OGLE EWS: The different types of variability present in 3084 OGLE-III light curves covering the 2003 to 2008 observing seasons.}
  \begin{tabular}{@{}cccccccc@{}}
   \hline
    Year 
          & Binary & Double & Unclassified & Other type & Finite Source & Point-Source\\
    & Lens & Source & Variable & of & events & Point-Lens\\
    & (high $q$) & & & Variability & excluded & (PSPL)\\          
   \hline
    2003 
          & 11 & 9 & 7 & 3 & 4 & 400 \\
    2004 
          & 16 & 10 & 7 & 4 & 4 & 536 \\
    2005 
          & 10 & 4 & 4 & 4 & 3 & 527 \\
    2006 
          & 12 & 7 & 3 & 7 & 5 & 511 \\
    2007 
          & 7 & 11 & 8 & 5 & 5 & 535 \\
    2008 
          & 11 & 12 & 9 & 4 & 4 & 575 \\
   \hline
    Total 
           & 67 & 53 & 38 & 27 & 25 & 3084 \\
   \hline
    As \% of sample 
                     & 2.03 & 1.61 & 1.15 & 0.82 & 0.76 & 93.62 \\
  \hline
  \label{tab:OGLE1}
  \end{tabular}
 \end{minipage}
\end{table*}

\citet{b25} published an estimate of the giant planet frequency beyond the ``snow line" using a selected sample of 13 very high magnification events in which they detect evidence of 6 planets. Even though the observations were not performed in a controlled fashion, the sample of events used in that analysis satisfied all the strict selection criteria and have such dense coverage that it can essentially be treated as a controlled experiment. Under the assumption that all planetary systems are Solar System ``analogs", they arrive at a first estimate of the frequency of solar-like systems of 16.7\%. Their sample included only M dwarf lenses with typical masses $\sim$0.5$M_{\odot}$.

In an independent analysis of 12 years of radial-velocity data for a sub-sample of 123 G and K stars, \citet{b36} looked for evidence of long period gas-giant planets at orbital distances of 3-6AU. After accounting for the efficiency of their survey and making the assumption of circular orbits, they concluded that no less than 3.3\% and no more than 37\% of stars in their sample host gas giant planets between 3 and 6AU.
\begin{figure*}
\includegraphics[angle=0,width = 0.75\textwidth]{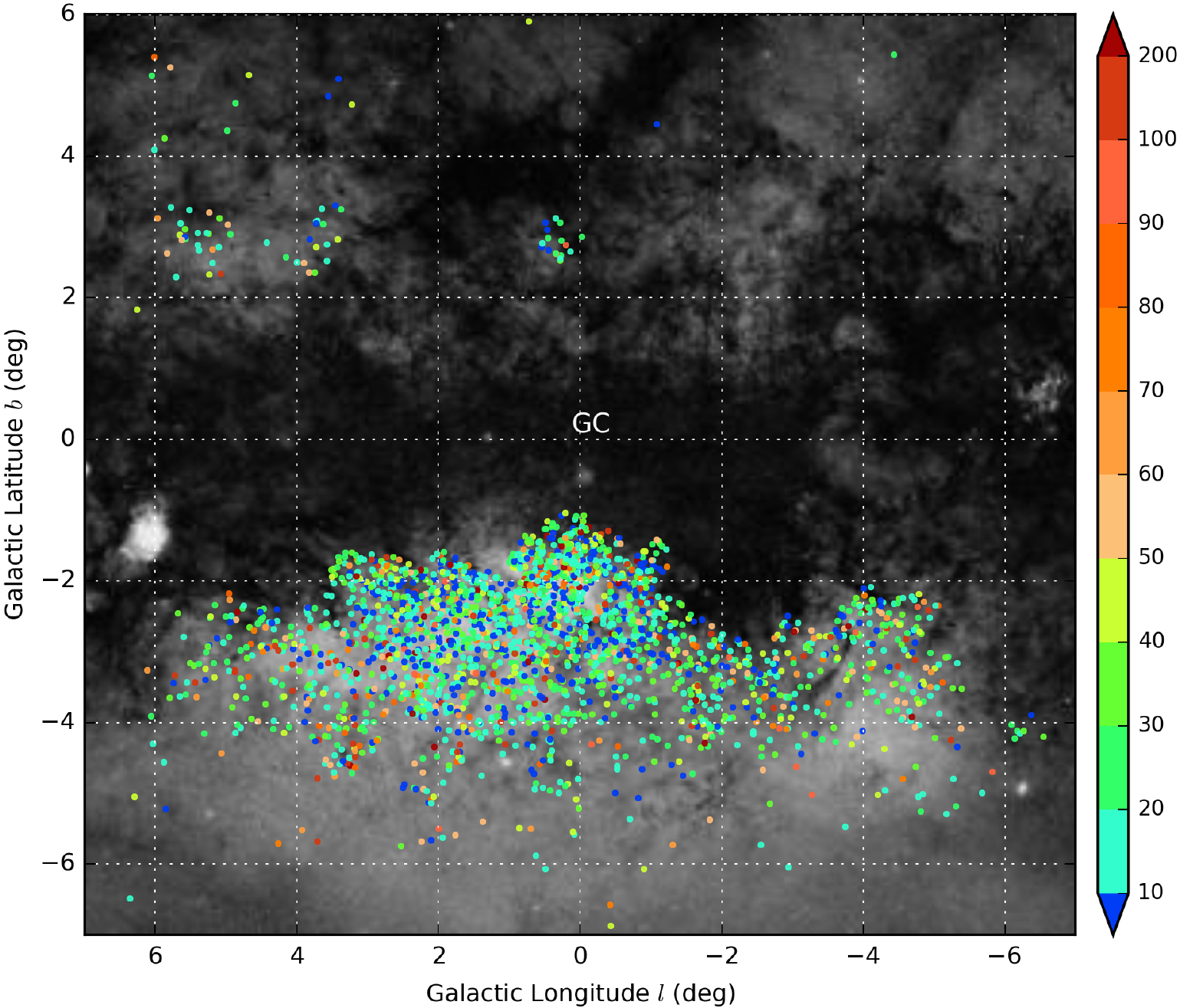}
\caption{Galactic longitude $l$ and latitude $b$ distribution of the sample of 2344 microlensing events. Events with shorter timescales are shown with bluer hues, while events with longer timescales are represented by redder hues. Event timescales are in days. The background is an optical image of the field.}
\label{gal_dist}
\end{figure*}

The Kepler team announced 1235 planetary candidates \citep{b45} after analyzing four months of observations and releasing the data on 155453 stars. Of these candidates, 74\% are smaller than Neptune and 54 were found in the temperature ranges corresponding to the habitable zone of their host stars. After correcting for selection biases, they report a 34\% frequency of candidate planets per star, with 17\% of stars hosting multiple planet candidates. Their second release \citep{b63}, based on sixteen months of data, yielded another 1091 planet candidates whose properties are similar to the previously published ones. However, they found evidence that smaller planets are more prevalent, 91\% of the new candidates had masses smaller than Neptune. The estimated fraction of stars with multiple planets had also increased from 17\% to 20\%.

The higher abundance of smaller planets relative to more massive ones is also corroborated by the work of \citet{b49}, who used 43 well sampled events extracted from six years of PLANET microlensing observations to place limits on the planetary abundance at distances of 0.5-10AU. They report that $17^{+6}_{-9}$\% of stars have planets with masses between 0.3-10$M_{Jup}$, whereas cool Neptunes ($m_{p}\sim$10-30$M_{\oplus}$) and super-Earths ($m_{p}\sim$5-10$M_{\oplus}$) are much more common with relative abundances of $52^{+22}_{-29}$\% and $62^{+35}_{-37}$\% respectively.

For the analysis presented in this work, we consider 3084 microlensing events from the OGLE-III survey, covering the years 2003 to 2008, after removing light curves dominated by non-planetary binary lens features ($\sim2\%$), double sources($\sim1.6\%$), contamination by variables or other types of unclassified variability ($\sim2\%$). We also exclude 25 events with clear finite source features that are incompatible with our simple PSPL model. We checked that the effects of ignoring finite source size for low magnification events are negligible, while the exclusion of a very small number of high magnification events only leads to a slight underestimation of the true detection efficiency (by $\sim 1-3\%$).
For an extensive discussion of variable and repeating events in the OGLE EWS\footnote{The OGLE Early Warning System announces microlensing events in progress.} during the period investigated here, see \citet{b26,b27,b28,b29,b30,b35,b67,b69}. 

The median cadence over all light curves in the sample, excluding gaps in the observations exceeding 30 days, is $\sim$1 observation per day, which offers good sensitivity to giant planets but only allows for weaker constraints to the presence of smaller-mass planets. Table \ref{tab:OGLE1} lists the annual breakdown of different types of variability announced by the OGLE EWS for the years considered. From left to right, the first column lists the year and subsequent columns list the number of i) binary lens candidate events where the less massive object is also of stellar mass, ii) events that can be attributed to double sources, iii) light curves of variable stars such as cataclysmic variables, iv) light curves that show more complex types of variability such as  microlensing of a variable star or other uncommon non-repeating types of variability, and v) number of events with clear finite source effects. Finally, the last column displays the number of events that are well fitted by a single point source, single point lens (PSPL) model.

\subsection[]{The Optical Gravitational Lensing Experiment}
Motivated by \citet{b31} and \citet{b32}, the OGLE survey started operations in 1992 with the aim of detecting microlensing events in the direction of the Galactic Bulge \citep{b33}. After carefully monitoring millions of stars, the first microlensing event was detected in 1993 \citep{b34} and new discoveries soon followed. The introduction of the Early Warning System (EWS) in 1994 \citep{b8} allowed newly detected microlensing events to be publicly announced in real time and heralded the era of follow-up observing campaigns. 

Constraints on the planet abundance based on an analysis of 145 OGLE-II events from the years 1998 to 2000 and, subsequently, from 321 events during the OGLE-III 2002 observing season have already been published \citep{b6,b7}. Here, we consider data from the third stage of the project, OGLE-III, and for the observing seasons 2003 to 2008. The OGLE survey observes in a controlled fashion so we do not use data collected by follow-up observations in our light curve analysis since these were obtained by observers reacting to alerts in an unpredictable manner and including such data would introduce a bias in our estimate of the detection efficiency. 

The datasets presented here are the latest photometric reductions of the OGLE-III images \citep{b68}. There are small differences with the photometry avalable on the EWS webpages which are mainly due to the use of different template images. We note that using either dataset for this analysis produces similar results.  
\begin{figure*}
\includegraphics[width = 1\textwidth]{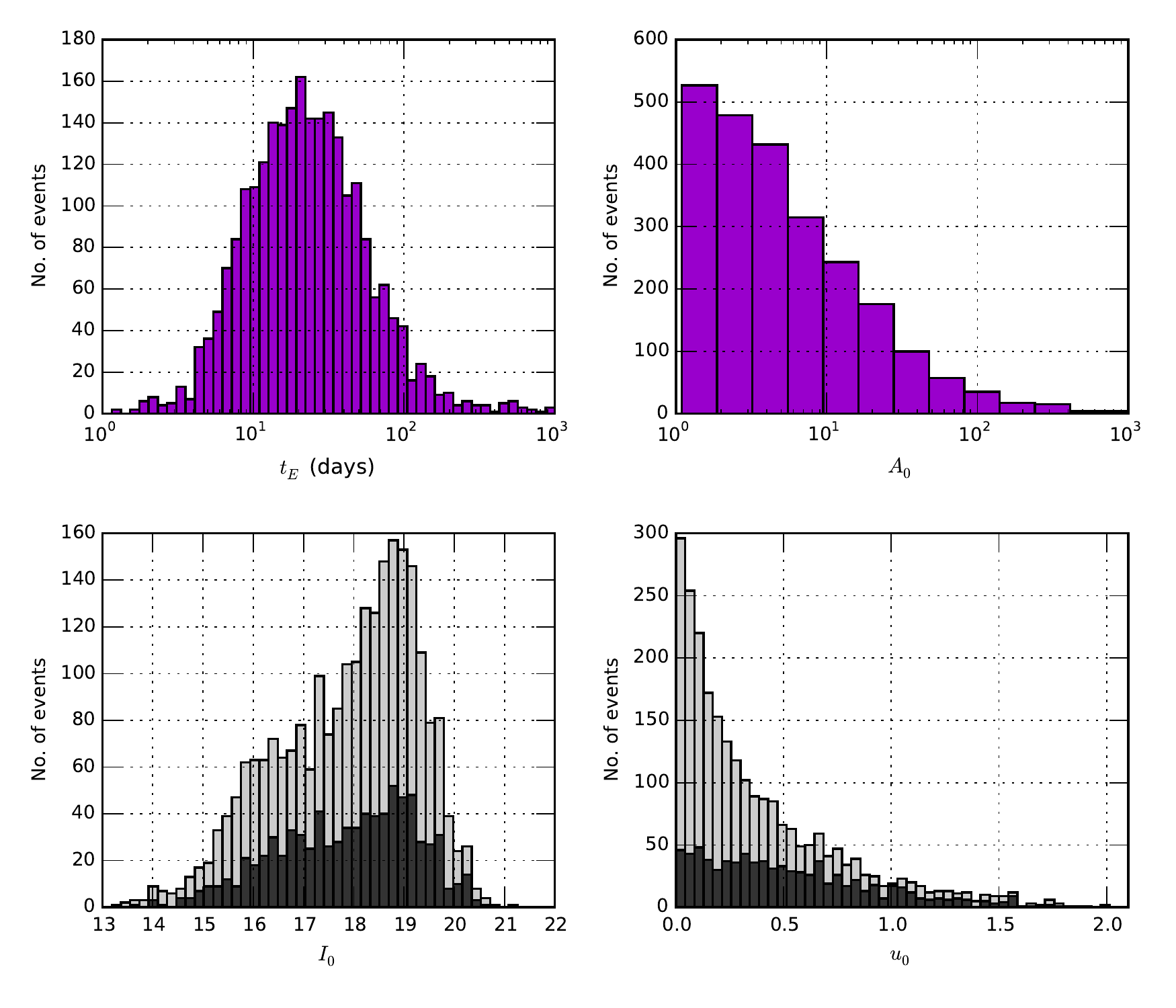}
\caption{{\bf Top left}: Distribution of the event timescale, $t_E$, for the sample of 2344 microlensing events. {\bf Top right}: Distribution of the maximum magnification, $A_0$. {\bf Bottom panels}: The darker histograms correspond to events that have a blend fraction $b\le$0.1 (Fig.~\ref{combined_2D_hist_all}-top right) while the lighter histograms are generated using all events in the sample. The panel on the left shows the distribution of the baseline magnitude, $I_0$, while the panel on the right shows the distribution of the source minimum impact parameter $u_0$. The distribution of $u_0$ is more uniform for less blended events.}
\label{combined_hist_all}
\end{figure*}

\subsection[]{Point-Source Point-Lens (PSPL) fitting}
The light curve produced by the simplest case of microlensing, that involving a single point lens and a single point source, can be fully characterized by four parameters: the event timescale $t_E$ (i.e. the time to cross $R_E$), the time of maximum magnification $t_0$, the baseline (unmagnified) magnitude of the source star $I_0$ and the source minimum impact parameter $u_0$\footnote{Or, equivalently, the maximum magnification $A_0$} i.e. the minimum source-lens separation, projected on the source plane, in units of $R_E$.

The magnification at time $t$ is
\begin{equation}
A(t) = \frac{u^2(t) + 2}{u(t) \sqrt{u^2(t) + 4}},
\label{eq_magnification}
\end{equation}
where 
\begin{equation}
u(t) = \left[u_0^2 + \left( \frac{t-t_0}{t_E} \right)^2\right]^{1/2}.
\end{equation}
The projected lens-source separation on the lens plane may be obtained from the magnification at any time from
\begin{equation}
u(t) = \left[\frac{2A(t)}{\sqrt{A^2(t)-1}} - 2\right]^{1/2}.
\end{equation}

\subsubsection{Accounting for blending}
Light curves obtained from the photometric analysis of observations of crowded fields, such as the Galactic Bulge, are commonly affected by blended light coming from stars near the lens or from the lens itself \citep{b38}. Blended light is added to the observed baseline flux of a microlensing event and can lead to incorrect estimates for the maximum magnification, $A_0$, and the event timescale, $t_E$, if unaccounted for.

Accounting for blending, the observed flux from the source star at time $t$ becomes $f(t) = f_s A(t) + f_b$ where $f_s$ and $f_b$ are the source and blend fluxes respectively and where $A(t)$ is given by Equation \ref{eq_magnification}. The observed magnification then becomes
\begin{equation}
A_{obs}(t) = \frac{f_s A(t) + f_b}{f_s + f_b} = \frac{A(t) + b}{1+b},
\end{equation}
where $b = f_b/f_s$ is the fifth parameter that we take into account.

For microlensing light curves that are densely sampled with good photometric quality, blending can be well constrained by the fitting process. Alternatively, it is also possible to resolve the stars contributing to the blended light by use of adaptive optics on large telescopes or from space \citep{b19,b40}. In cases where the light curve is not finely sampled, blending is only loosely constrained and the fitting algorithms can converge to local minima around the seed value for the blending parameter \citep{b39}.

To ensure that we converge on a reasonable value for the blending parameter, we perform a Bayesian blend analysis before the actual fits. We set up a grid of 51 blend values in $log(b)$ and search for the best solution which we then use as a starting point for our subsequent fits. The grid is uniform in $log(b)$ and ranges from -2 to 2.

\subsubsection{Treatment of the error bars}
\label{treatment_errbars}
In addition to these five parameters, we introduce a sixth; an additive flux error that readjusts the reported size of the error-bars. This parameter accounts for the observed scatter in the measurements of the unlensed flux\footnote{If this parameter is not included in the fits, the residuals are larger than expected on the unlensed part of the light curves.}. The error bars associated with the flux measurements are $s_i = \sqrt{{\sigma_0}^2 + {\sigma_i}^2}$ where ${\sigma_0}^2$ is the variance of an additive flux error and $\sigma_i$ is the {\it flux} error bar corresponding to the originally reported magnitude error bar on the $i^{\rm th}$ photometric measurement.
\begin{figure*}
\includegraphics[width = 1\textwidth]{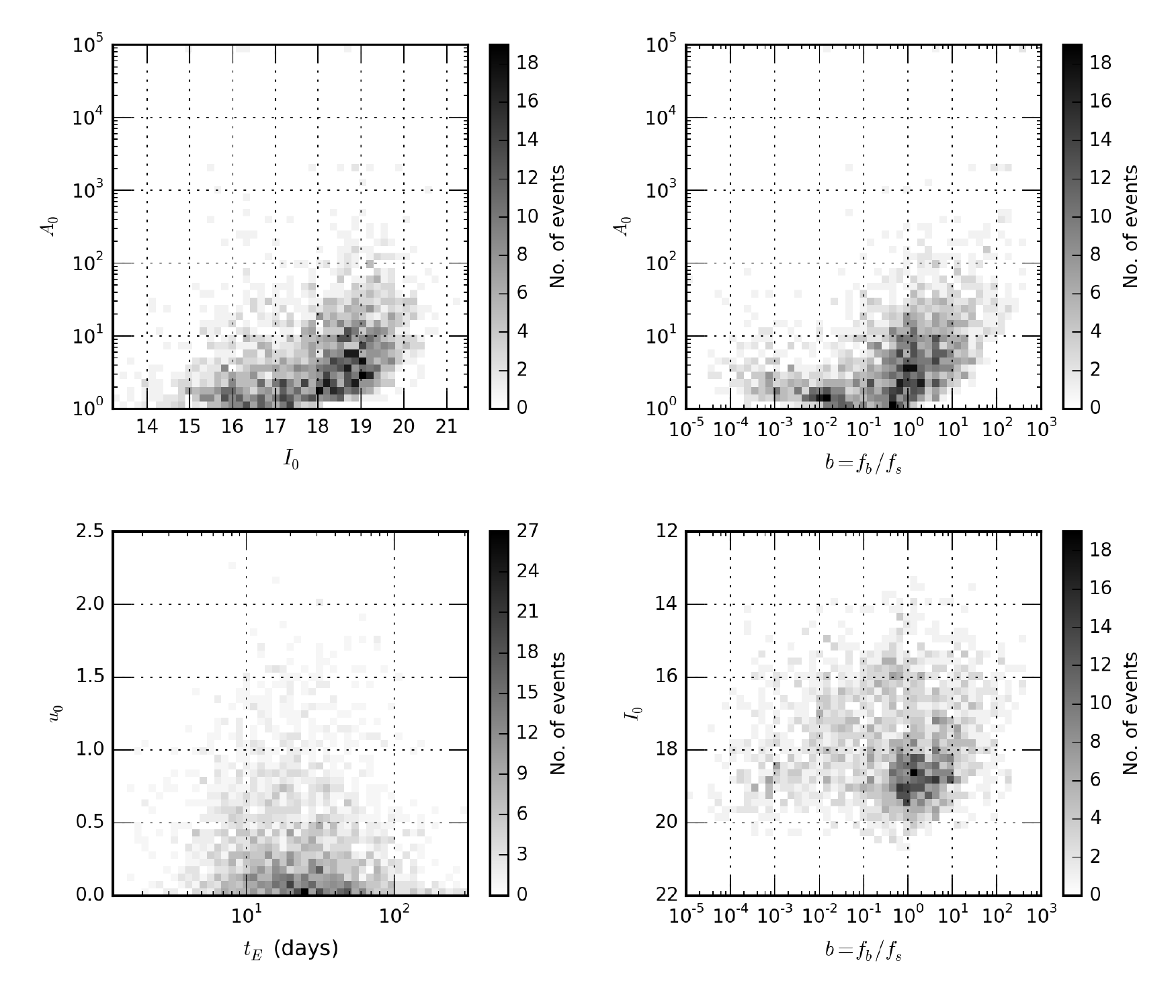}
\caption{2D histograms. {\bf Top left}: Maximum magnification as a function of the baseline magnitude. Fainter stars are detected when they are more highly magnified. {\bf Top right}: Maximum magnification as a function of the blend fraction. {\bf Bottom left}: Minimum impact parameter as a function of the event timescale. {\bf Bottom right}: Baseline magnitude as a function of the blend fraction.}
\label{combined_2D_hist_all}
\end{figure*}

A blind parameter search can sometimes converge towards unphysical solutions. Therefore we incorporate prior distributions on the parameter space\footnote{See Appendix A for a description of the priors used.} and perform a Bayesian parameter estimation similar to \cite{b64}.

Using Bayes' theorem, we can write the posterior probability distribution over the model parameters $\phi$ as a function of the data $D$
\begin{equation}
P(\phi|D) = \frac{P(D|\phi)P(\phi)}{\int{P(D|\phi)P(\phi)d\phi}},
\label{bayes}
\end{equation}
where $P(\phi)$ is the prior probability distribution of the parameters and $P(D|\phi)$ is the likelihood function. The denominator ensures that 
$P(\phi|D)$ is normalized as a probability distribution over the parameters. We want to maximize the posterior probability of the model, $P(\phi|D)$.

Assuming $N$ data points with associated independent Gaussian errors $s_i$, we may now write the likelihood of our model parameters $\phi$ as
\begin{equation}
L(\phi) = P(D|\phi) = \frac{e^{-\chi^2 / 2}}{{\displaystyle \prod_{i=1}^N} (2 \pi s_i^2)^{1/2} }.
\end{equation}
Taking into account the priors, we may write
\begin{equation}
-2\ln [L \times P(\phi)] = \chi^2 + 2 \sum_{i=1}^{N} \ln s_i - 2 \ln P(\phi) + N \ln 2\pi.
\label{max_lik2}
\end{equation}
The last term in Equation~\ref{max_lik2} is a contant that can be ignored during minimization.

We perform initial PSPL fits to all microlensing light curves in our sample by adjusting these six parameters using a downhill simplex algorithm which minimizes Equation~\ref{max_lik2}. Note that the sample includes five published events with known planetary anomalies\footnote{OB-03-235 \citep{b70}, OB-05-071 \citep{b71}, OB-05-390 \citep{b72}, OB-06-109 \citep{b73}, OB-07-378 \citep{b74}} but the OGLE-III dataset alone is not sufficient for full characterization and the PSPL fits adequately describe the overall shape of the light curve.

\subsection{Fitting results}
We fit a PSPL model to 3084 microlensing light curves which correspond to microlensing events that were detected by the OGLE-III survey and announced through their Early Warning System between the years 2003 and 2008. Selection criteria based on event light curve quality were applied (see section \ref{og_candi}) and were satisfied by 2433 light curves. The sample is large enough that it allows us to explore the distributions of the most interesting parameters and the correlations between them. 

The distribution of events in Galactic longitude and latitude is shown in Fig.~\ref{gal_dist}. Events with shorter timescales are identified by bluer hues and longer events are shown redder. The regions closer to the Galactic Centre are dominated by shorter events, whereas events with longer timescales are located predominantly in the outermost areas. For a comprehensive comparison of the timescale distribution of OGLE-III events with the most recent Galactic models we refer the interested reader to \cite{b75}.

In Fig.~\ref{combined_hist_all} (top right) we plot a histogram of the distribution of the maximum magnification for the sample of 2433 light curves. This parameter ranges from as low as 1.06 to values of above 1000 for a handful of events. In extreme cases the fits may begin converging towards very high magnification values due to sampling gaps around the peak and the occasional outlier. To safeguard against this, we impose a limit $A_0 \le 10^5$ on the maximum magnification during the fitting process.

The top left panel of Fig.~\ref{combined_hist_all} shows the distribution of the event timescales, i.e. the time it takes for the source to traverse a distance equal to the Einstein ring radius of the lens. This peaks at $\sim$20 days with a tail of long event timescales, more than 100 days, that may be caused by more massive lenses or more distant lenses or closer sources.

In the bottom left panel of Fig.~\ref{combined_hist_all} we plot the distribution of the $I$-band baseline magnitude for our sample. This peaks at $I_0\sim$19, beyond which the sensitivity quickly drops unless the source star is highly magnified. The darker histogram corresponds to events that have a fitted blend fraction value $b\le$0.1 (also see the top right panel in Fig.~\ref{combined_2D_hist_all}) whereas the lighter histogram is generated using all events in the sample. As expected, there is a larger number of fainter stars that are more highly blended. 
\begin{figure}
\includegraphics[width = 0.5\textwidth]{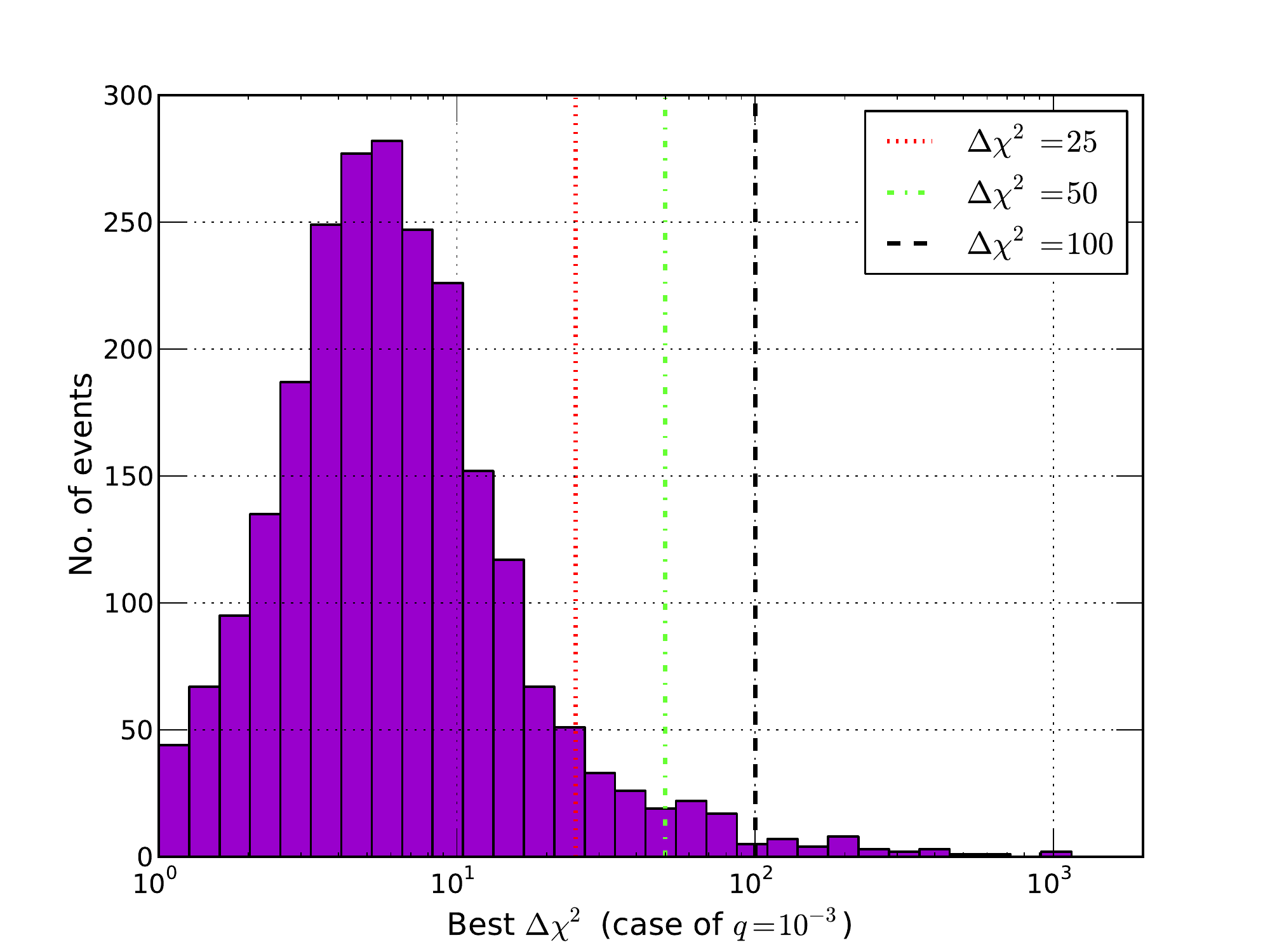}
\caption{Distribution of the best ${\Delta\chi^2}$ (highest value in detection map) for the sample of 2433 light curves generated for a mass ratio $q=m_p/m_*= 10^{-3}$. The detection maps are generated for three threshold values which are marked with the dotted (red), dot-dashed (green) and dashed (black) lines.}
\label{delta_chi_plot}
\end{figure}

Of particular interest is the distribution of the minimum impact parameter, $u_0$, displayed in the bottom right panel of Fig.~\ref{combined_hist_all}. Events with a blend fraction $b\le$0.1 are represented by the darker histogram, while the lighter histogram is produced from the entire sample. As pointed out by \citet{b50}, the observed distribution is non-uniform, but that is merely a selection effect which favors the detection of faint, more blended, events when they are more highly magnified. The distribution of $u_0$ is more uniform for less blended events, as expected.

It is instructive to consider how the fitted baseline magnitude correlates with the maximum magnification $A_0$. This is shown in the top left panel of Fig.~\ref{combined_2D_hist_all} where selection effects are apparent. Fainter events are detected when they are more highly magnified.

The top right panel of Fig.~\ref{combined_2D_hist_all} shows the fitted blend fraction versus the maximum magnification. The distribution shows two almost distinct populations, one peaking at $b\sim 10^{-2}$ and another peaking at $b\sim 1$. The less blended sources on the left side are less magnified than the more blended sources on the right. The less blended sources have lower peak magnifications ($A_0\sim$2) while more blended sources have $A_0\sim$5. One reason why this occurs is because the fitting process attempts to compensate for light curves with bad sampling at the peak and/or wings as well as single outliers at the peak by increasing the blend fraction and biasing the fitted magnification to higher values. As the blend fraction and magnification are correlated quantities, greater uncertainty in one translates to greater uncertainty in the other, so events with higher and more uncertain $A_0$ values will also have higher and more uncertain $b$ values.

The distribution of impact parameters as a function of the event timescales is shown in the bottom left panel of Fig.~\ref{combined_2D_hist_all}. As in the bottom right panel of Fig.~\ref{combined_hist_all}, we see a preponderance of events with smaller impact parameters.

In Fig.~\ref{combined_2D_hist_all} (bottom right), we plot the blend fraction as a function of the baseline magnitude. The majority of events cluster around $b\sim$1 in agreement with the analysis presented in \citet{b58}.
\begin{figure}
\includegraphics[width = 0.5\textwidth]{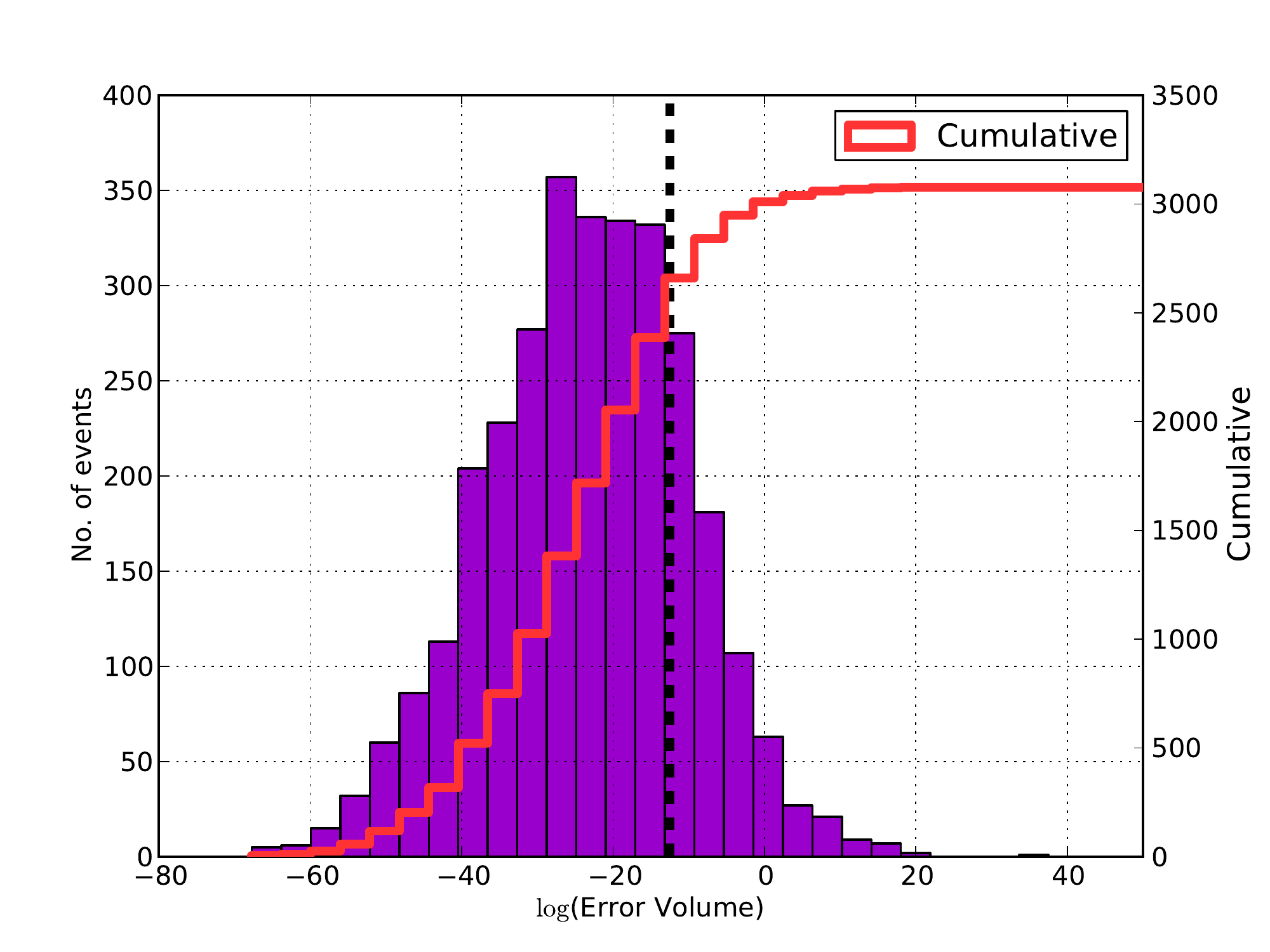}
\caption{Histogram of the distribution of the {\it error volume} obtained by the computing the hyper ellipsoid of the uncertainties based on the Eigenvalues of the Fisher matrix. Our selection threshold is indicated by the dashed black line and all events with $\log$({\it error volume}) $>$ -12.5 are rejected.}
\label{fischer}
\end{figure}

\section{Search for low mass companions}
\subsection{$\Delta\chi^2$ detection maps}
For a given planet-to-star mass ratio $q$, we set up a fine grid of planet positions on the lens plane ($x$,$y$) and fit a static binary lens model to the data at each of those locations \citep{b6,b7}. This grid search must be fine enough so that no viable models are missed and therefore we conservatively choose a step-size of $\sqrt{q}/4$ for the grid. Using the binary lens evaluation at each grid position and the previous PSPL fit, we construct a $\Delta\chi^2 = \chi_{\mbox{\tiny single}}^2 - \chi_{\mbox{\tiny binary}}^2$ detection map for each event, where $\chi_{\mbox{\tiny single}}^2$ and $\chi_{\mbox{\tiny binary}}^2$ are the minimum $\chi^2$ values of the point-source, point-lens and point-source, binary-lens models respectively.

We define the {\it detection zone} as a region on the lens plane where a light curve anomaly is
confirmed by the observations, that is, it exceeds a given $\Delta\chi^2$ threshold value. This threshold must be set high enough so that the rate of false detections is minimized but also low enough so that possible detections are not completely suppressed. We generate maps for three different ${\Delta\chi^2}_T$ threshold values: 25, 50 and 100. The middle panels of Fig.~\ref{plens_plots} present two examples of such maps which were generated for a mass ratio of $q=m_p/m_*= 10^{-3}$ and a threshold ${\Delta\chi^2}_T$=100. 

The histogram displayed in Fig.~\ref{delta_chi_plot} shows the distribution of the highest ${\Delta\chi^2}$ value, extracted from the detection map of each event, for the sample of 2433 light curves and the three threshold values.

\subsection{Planet detection probability}
\label{pdp}
For each event we calculate the planet detection probability for ten different mass ratios, $10^{-2}$ to $10^{-5}$. The detection probability for a planet of mass ratio $q$ at projected position $(x,y)$ on the lens plane for a specific orbital radius $\alpha$ is given by
\begin{equation}
P(\mbox{det}|\alpha,q) = \int P(\mbox{det}|x,y,q) P(x,y|\alpha) dx dy.
\label{det_prob}
\end{equation}
The first term in the integral above is
\begin{equation}
P(\mbox{det}|x,y,q) = 
\begin{cases}
1& \text{if  $\Delta\chi^{2} > {\Delta\chi^2}_{T}$}, \\
0& \text{otherwise}, \\
\end{cases}
\end{equation}
and it becomes significant when the planet located at position $(x,y)$ happens to perturb one of the images of the source generated at the times of the observations corresponding to the data points of the event light curve. The second term, $P(x,y|\alpha)$, is obtained by assuming a circular orbit of radius $\alpha$ for the planet, drawing a random orientation for the orbital plane from a uniform distribution over the surface of a sphere, and projecting it on the lens plane at $(x,y)$. This generates a radially symmetric distribution centered on the lens which increases as $(d/\alpha)^2$ and peaks at $d=\alpha$, beyond which the probability is 0. We may write this term as
\begin{equation}
P(x,y|\alpha) =
\begin{cases}
\frac{1}{2 \pi \alpha \sqrt{\alpha^2-d^2}} & \text{for ${d = \sqrt{x^2 + y^2} < \alpha}$}, \\
0 & \text{otherwise}. \\
\end{cases}
\end{equation}
This means that the detection probability given by equation (\ref{det_prob}) is the result of summing up the fraction of the time that a planet with an orbit of radius $\alpha$ spends inside the {\it detection zones}. A planet's presence is inferred by perturbations caused to one of the images of the source which appear around the Einstein ring of the lens. Consequently the strongest {\it detection zones} are also located around the Einstein ring 
and the highest detection probability is at $\alpha \simeq R_E$.
\begin{figure*}
   \centering
   \begin{tabular}{cc}
      \includegraphics[angle=0,width=0.51\textwidth]{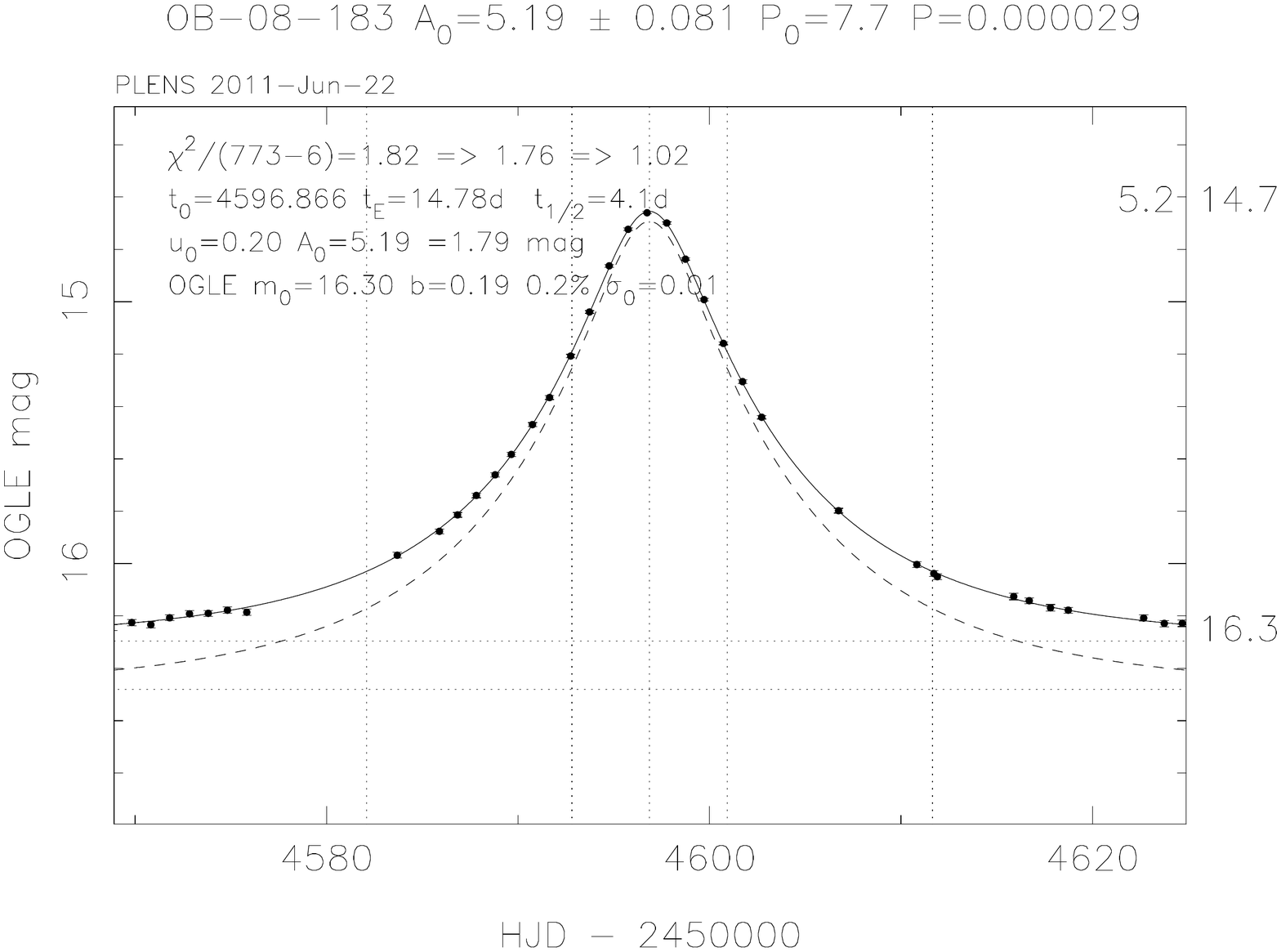}&
      \includegraphics[angle=0,width=0.51\textwidth]{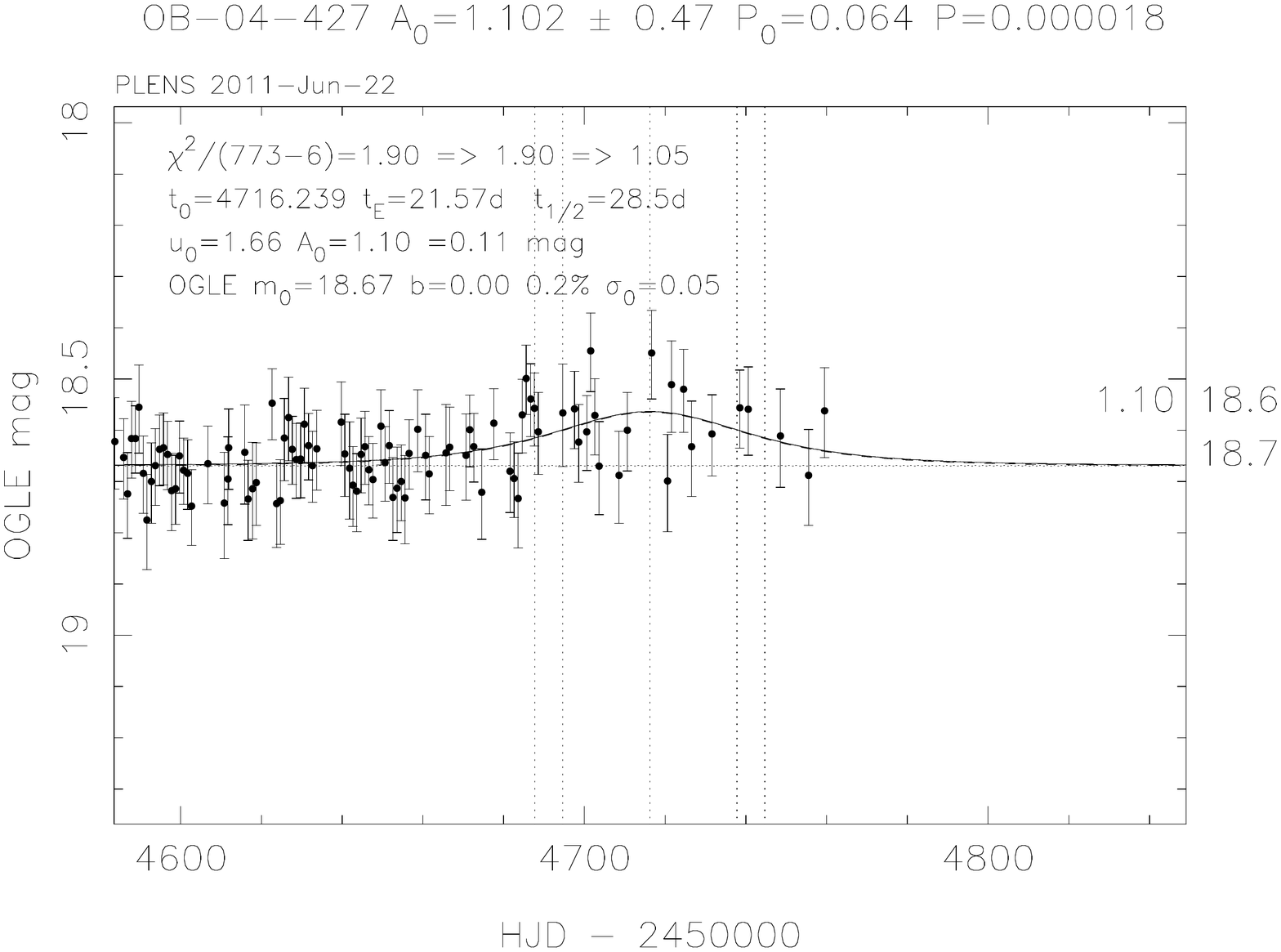}\\
      \includegraphics[angle=0,width=0.51\textwidth]{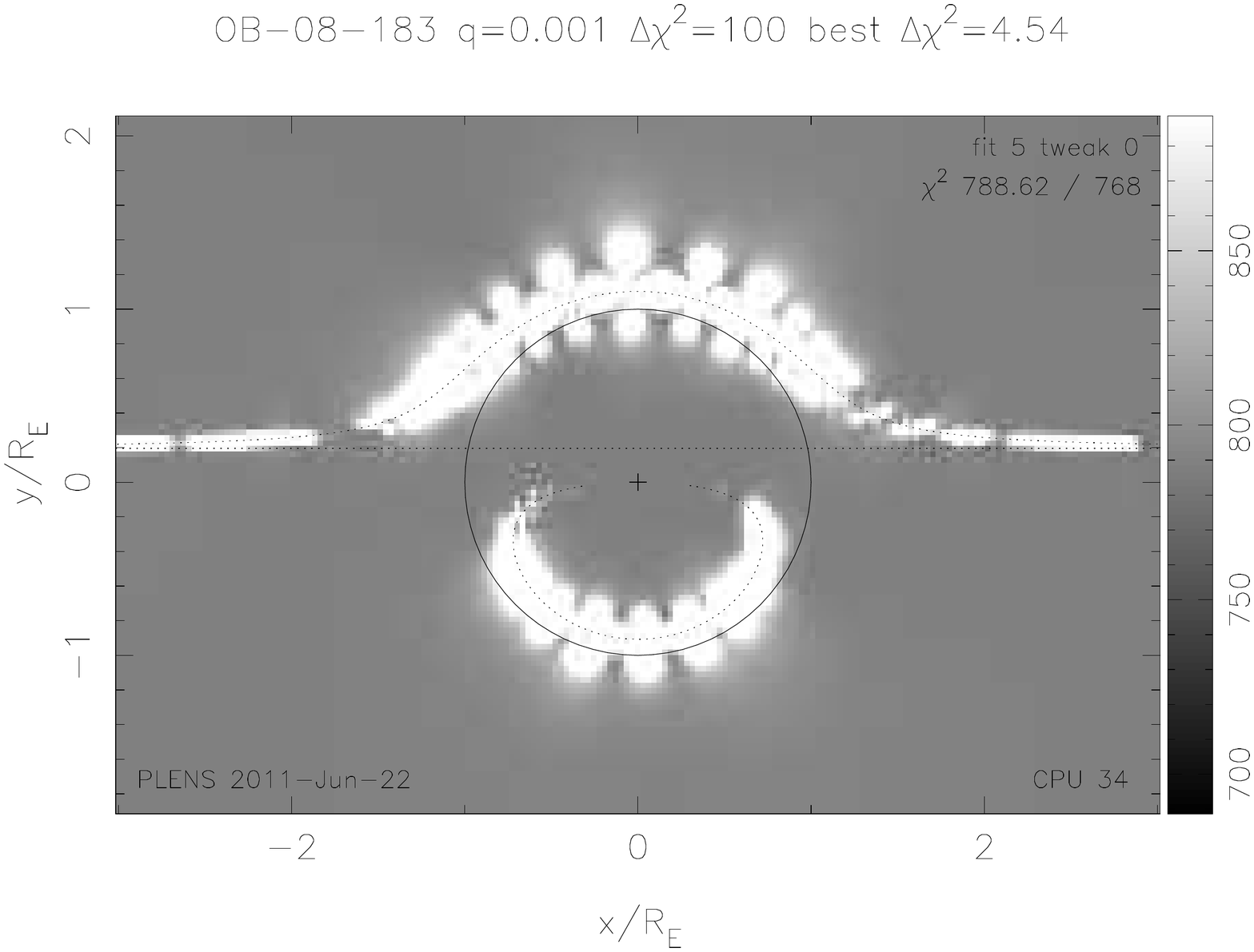}&
      \includegraphics[angle=0,width=0.51\textwidth]{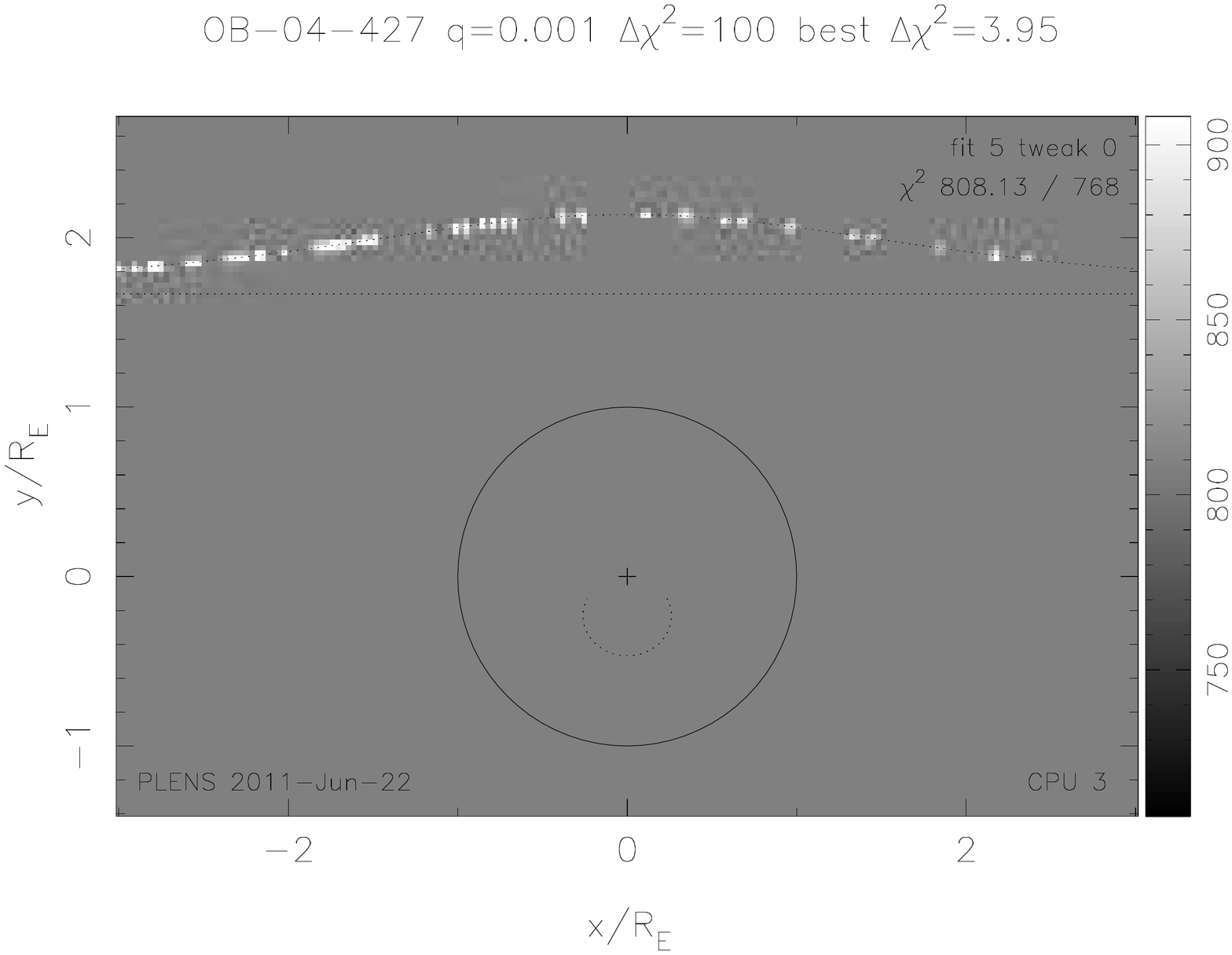}\\
      \includegraphics[angle=0,width=0.51\textwidth]{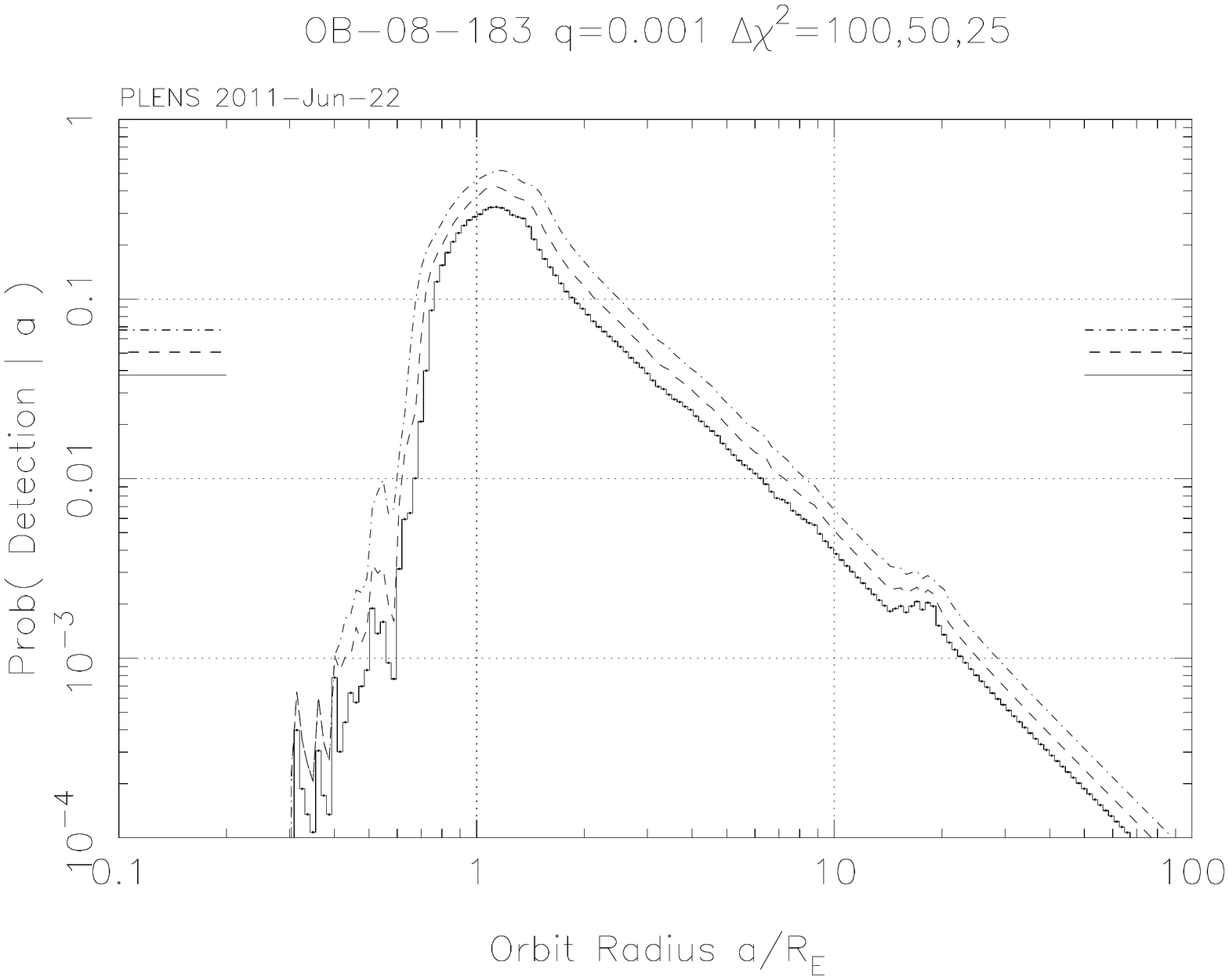}&
      \includegraphics[angle=0,width=0.51\textwidth]{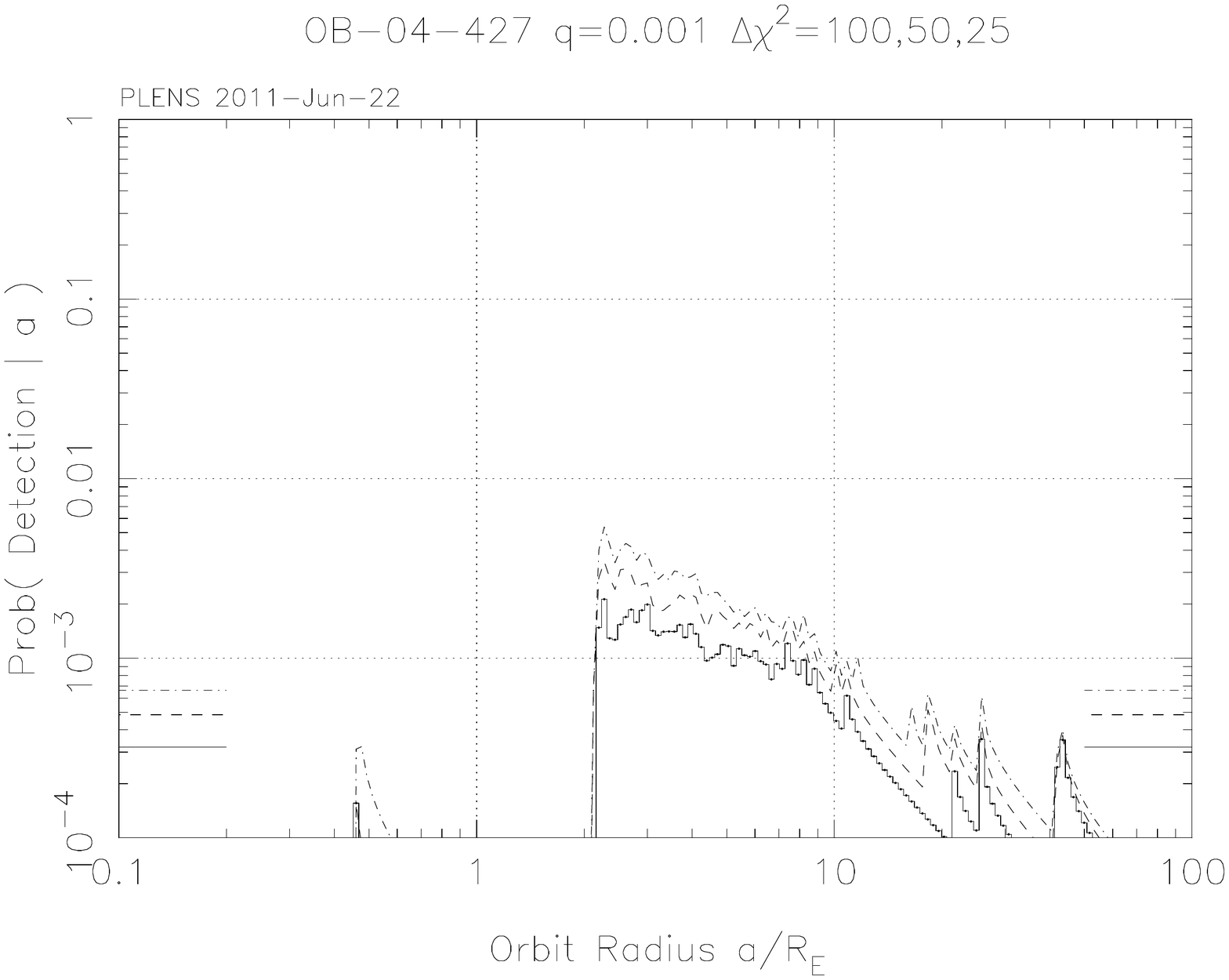}\\
   \end{tabular}
\caption{Top: Best PSPL fit to the light curves of OGLE-2008-BLG-183 and OGLE-2004-BLG-427. The solid curve shows the best-fit PSPL model including blending whereas the dashed curve shows the unblended light curve. The normalized residuals are shown below the fitted light curves. Middle: The corresponding $\Delta\chi^2$ detection maps for these two events. The white zones mark regions where the presence of a planet of mass ratio ($q=10^{-3}$) can be excluded at $\Delta\chi^2$=100 given the data. Bottom: Detection probability at different orbital radii for a planet with mass ratio $q=10^{-3}$ for these two events. The three horizontal lines at the edges of the plot mark the detection probability if planets are uniformly distributed across the range of the plot. From bottom to top, the curves are for threshold values $\Delta\chi^2 >$ 100, 50, 25 respectively.}
\label{plens_plots}
\end{figure*}

To illustrate the methodology we provide representative examples of two extreme event cases, namely OGLE-2008-BLG-183 and OGLE-2004-BLG-427, whose light curves are shown in the top two panels of Fig.~\ref{plens_plots}. The associated detection zones are shown in the middle panels. In the case of OGLE-2008-BLG-183, the detection probability of a giant planet at $\sim$1$R_E$ is of the order of 30\% (for a significance level of $\Delta\chi^2 >$ 100), as shown in the bottom panels of the figure. For OGLE-2004-BLG-427 on the other hand, the detection probability in the same parameter range does not exceed 0.2\%.

\subsection{Assessment of the OGLE light curves}
\label{og_candi}

\subsubsection{Treatment of the sample}
Not all light curves are useful for our analysis. Our sample contains light curves that we cannot fit with sufficient accuracy and where the PSPL parameters are only loosely constrained. These need to be identified and removed from our sample. To that effect, we define and calculate the {\it information content} of each light curve and use this as our criterion for selection. 

Following the classical approach introduced by \citep{b48}, the information content is estimated from the sensitivity of the log-likelihood with respect to the event parameters. For an observation of likelihood $L$ and an associated parameter vector $p_i$, the Fisher matrix is defined as:
\begin{equation}
F_{ij} = \left\langle \left(\frac{\partial \log(L)}{\partial p_i}\right) \left(\frac{\partial \log(L)}{\partial p_j}\right) \right\rangle
\end{equation}
where the expected value is an ensemble mean over all possible light curve realizations given a fiducial model. It depends exclusively on sampling and reported uncertainties but not on the brightness measurements themselves. The reported uncertainties are subject to the Cram\'{e}r-Rao bound
\begin{equation}
\textrm{cov}_{ij} \geq \left(F^{-1}\right)_{ij},
\end{equation}
which rejects only those events that are theoretically insufficient for characterizing the light curve and includes the uncertainties and theoretical correlations of all parameters at the same time. The total information content of the Fisher matrix is determined for all observations by calculating the {\it error volume} in parameter space units which is the hyper-volume of the multidimensional ellipsoid of the covariance matrix. By definition, this hyper-volume is proportional to the product of the eigenvectors of the covariance matrix. 

We set the selection threshold for retaining a light curve to $\log$({\it error volume}) $<$ -12.5, which corresponds to the product of the median variances for each fitted parameter in our data set. Our results are plotted in Fig.~\ref{fischer} and our selection threshold, indicated by the dashed black line, leads to the rejection of 651 events as unsuitable for further analysis (negligible information content). This leaves 2433 light curves in our sample. 

Referring back to the example event cases we presented at the end of section \ref{pdp}, OGLE-2008-BLG-183 is among the best 5\% and survives the selection, whereas OGLE-2004-BLG-427 belongs to the worst 5\% and gets rejected.

\subsubsection{Noise properties of the data}
\label{thresh}
\begin{figure}
\includegraphics[width = 0.5\textwidth]{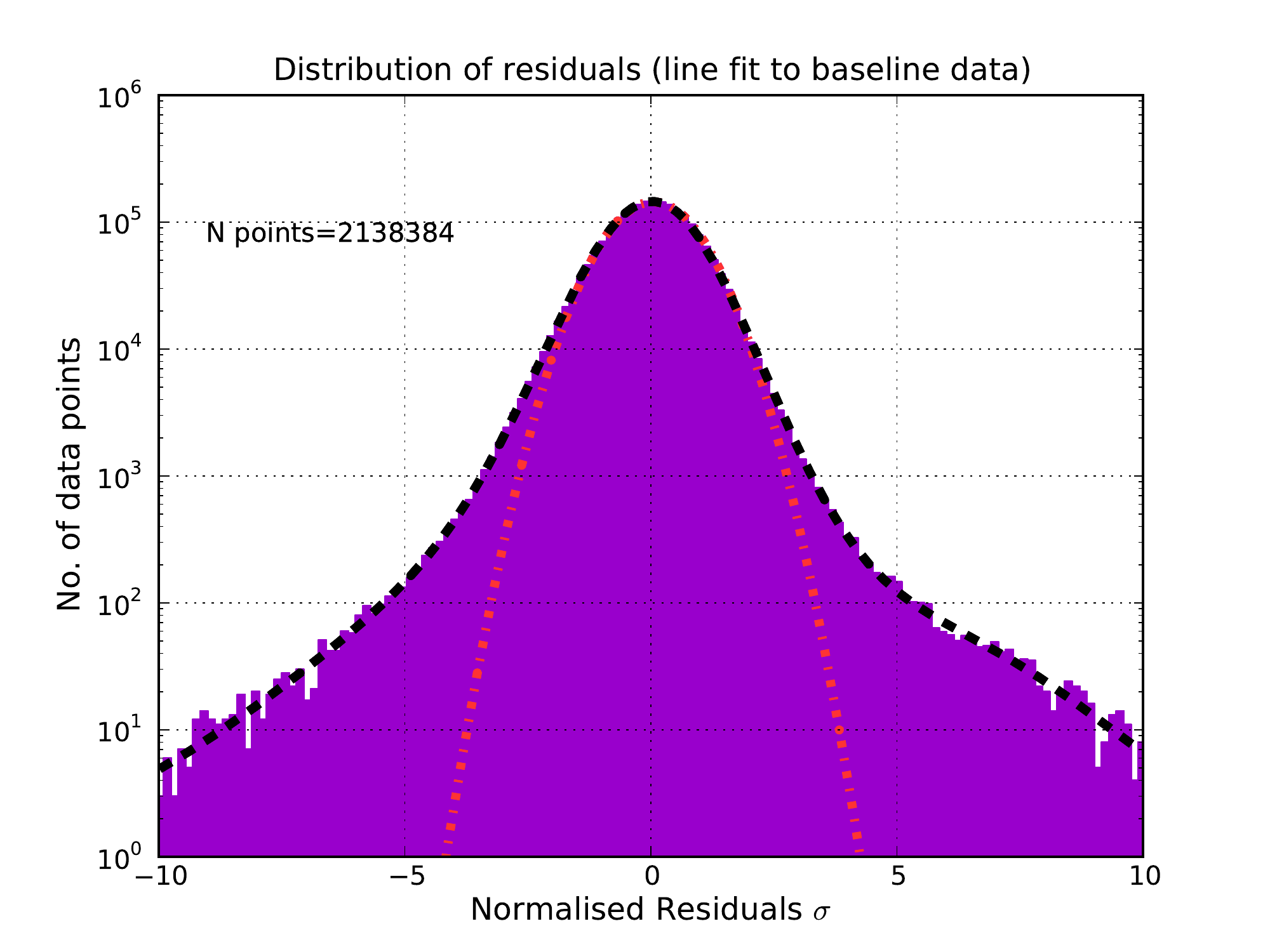}
\caption{Histogram of the distribution of the normalized residuals of straight line fits to the baseline data of every remaining event in the sample. The tails of the distribution are broader than expected from a purely Gaussian model (red dot-dashed line). The model that best fits the observed distribution is a combination of three Gaussians (black dashed line).}
\label{baseres}
\end{figure}

It is commonly assumed that the reported uncertainties in the {untreated} data are normally distributed. We test this assumption by discarding data taken during the microlensing phase and fitting a constant flux to the baseline data for all events, after having converted magnitudes back to fluxes. Fig.~\ref{baseres} shows the resulting histogram distribution of the residuals. The tails of this distribution are broader than expected by a purely Gaussian distribution (red dot-dashed line) and a Kolmogorov-Smirnov test rejects this hypothesis. An empirical fit to the distribution is achieved using the sum of three Gaussians 
\begin{equation}
\Phi(x) = \sum_{i=1}^3 a_i e^{-\frac{(x-b_i)^2}{2c_i^2}}
\label{3gauss}
\end{equation}
fitted to $\log_{10}(N_d)$ (where $N_d$ is the total number of data points at each $\sigma$ bin) as indicated by the black dashed line on the plot. The values of the coefficients are given in table~\ref{tab:valcoeff}.

{Fig.~\ref{baseres} illustrates that the reported error-bars of the raw OGLE lightcurves do not represent the true photometric uncertainties. Since outlying data points might be due to genuine anomalies, we avoid rejecting them. However, in order to account for this underestimate of the true errors, all error-bars are rescaled during the fitting process as already described in section~\ref{treatment_errbars}.}

\subsection{Estimating the survey sensitivity to planets}
\begin{figure}
\includegraphics[width = 0.5\textwidth]{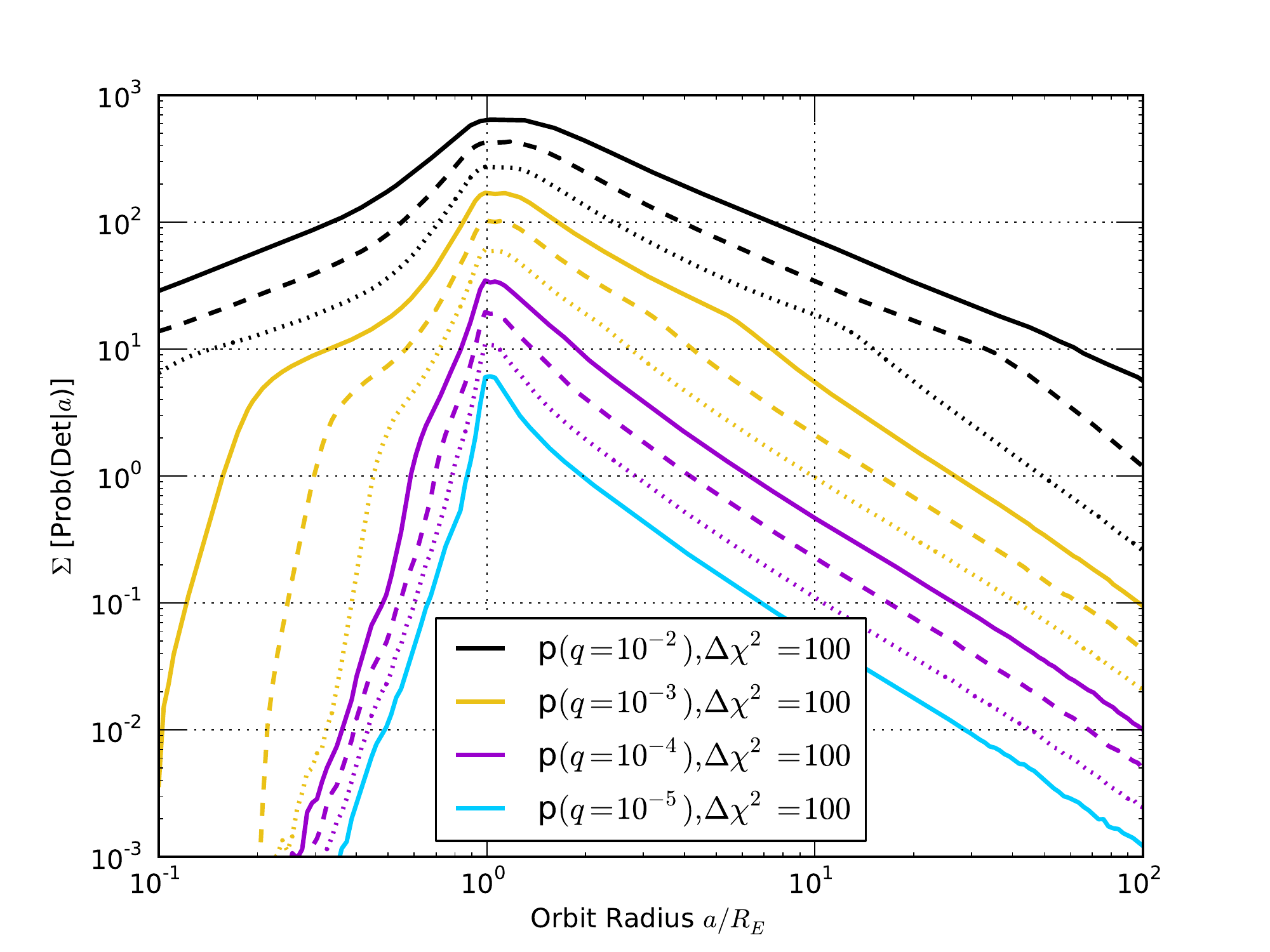}
\caption{Expected number of detections as a function of orbital radius in units of $R_E$ based on the analysis of 2433 OGLE-III microlensing events. We assume that each star has a planet of the specified mass ratio $q$ at each value of the orbital radius $\alpha$. The graph presents the results for ten different mass ratios, $q=10^{-2}$(top curve) to $10^{-5}$(bottom curve), equidistant in log space. We only plot the values corresponding to our selected threshold of ${\Delta\chi^2}_{T}$ = 100.}
\label{combined_prob}
\end{figure}

Fig.~\ref{combined_prob} shows the expected number of detections plot\footnote{Note that this plot is almost identical whether we use the entire original sample of 3084 light curves or the cleaner sample of 2433. This is not surprising since the rejected light curves offer no sensitivity to planets and hence contribute virtually nothing to the final sum.}, obtained by summing up the detection probabilities over all stars $i$ in the sample, for a specific value of the mass ratio, $q$, and at each value of the orbital radius $\alpha$:
\begin{equation}
P(\alpha,q) = \sum_i P_i(\mbox{det}|\alpha,q).
\end{equation}

We have performed this calculation for ten values of the mass ratio, from $q=10^{-2}$ to $q=10^{-5}$ taking equal steps in log space and considering three different thresholds ${\Delta\chi^2}_{T}=25, 50, 100$.

If all stars have $n_p$ planets of mass ratio $q$ orbiting them at orbit radius $\alpha$, then we expect $<n_d> = n_p P(\alpha,q)$ detections \citep{b6,b7}. From Fig.~\ref{combined_prob}, the highest value for the expected number of detections is obtained, as expected, for orbital radii close to the Einstein ring radius, $\alpha \approx R_E$. The expected number of detections drops rapidly for planets that are located deeper inside the Einstein ring and for planets that are much further out. The expected number of detection remains significant from $\alpha\sim$0.6 to $\sim$10 and decreases by a factor of ten for a factor of ten drop in the mass ratio, $q$.

\subsection{Using a Galactic Model}
Fig.~\ref{combined_prob} presents our results in terms of companion-lens mass ratio $q$ and companion-lens projected separation $a$ in units of the Einstein ring radius of the lens. In order to convert our distribution from dimensionless units $a$ and $q$ to physical units $\alpha/$AU and $m_{\oplus}$, we employ a Galactic Model. A detailed description of our model is beyond the scope of this paper, but we discuss below the basic assumptions leading to the $\log m_L$-$\log t_{E}$-$\log R_{E}$ relation that we use for interpreting our results. 

\begin{figure}
\includegraphics[width = 0.5\textwidth]{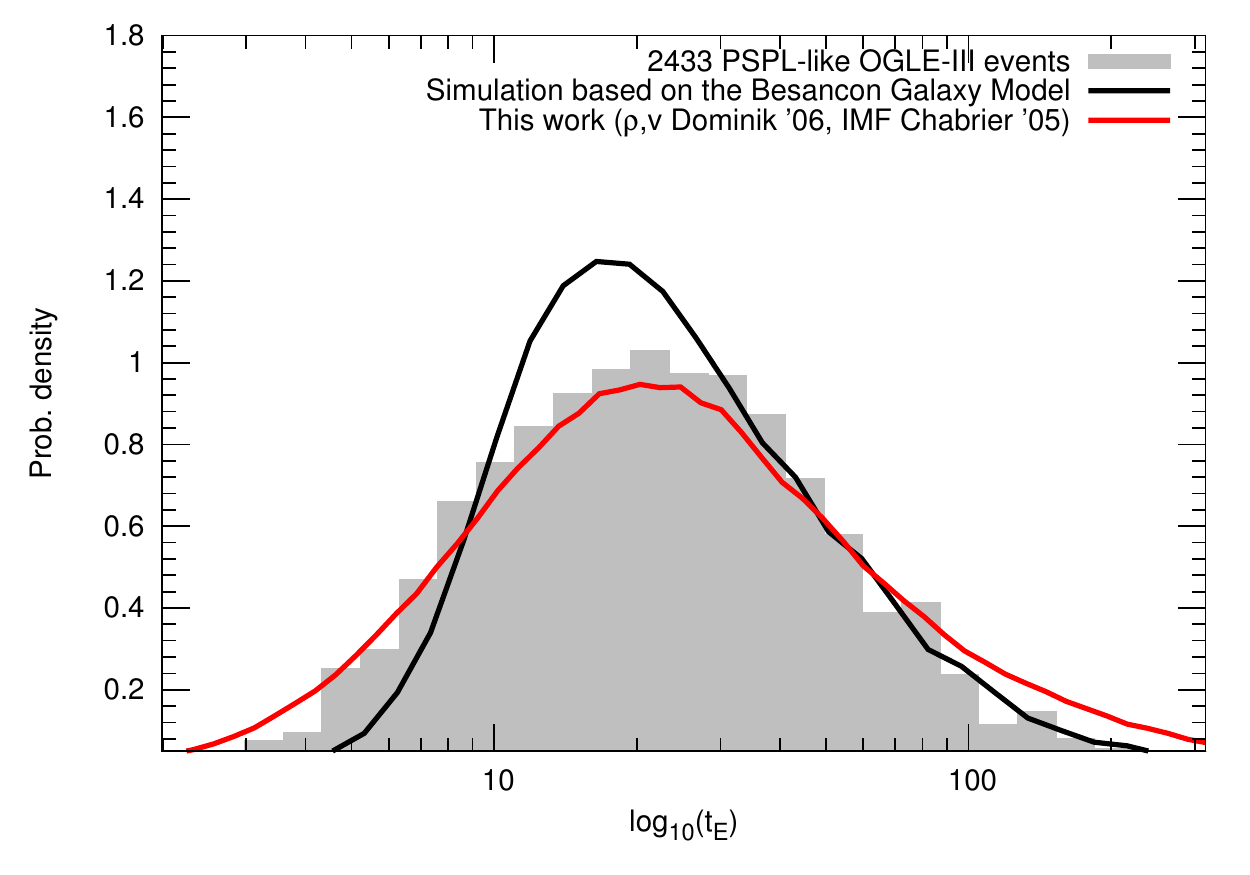}
\caption{Predicted timescale distribution obtained from our selection of 2433 PSPL events and the corresponding distributions from the Besan\c{c}on model compared to our Galactic Model.}
\label{galmodel_comparison}
\end{figure}

For each set of Galactic longitude $l$ and latitude $b$, corresponding to the location of each event in the Galactic Bulge, we use our Galactic Model to sample the distribution of relative distances, velocities and masses, assuming the distribution of lens masses follows the \citet{b56} initial mass function (IMF). Lens-source relative proper motions are dominated by velocity dispersion. The mass density of lenses in the Galaxy follows \citet{b57}. This choice of parameters aims to reproduce the observed timescale distribution in units of $t_E$.
\begin{table}
\centering
 \begin{minipage}{140mm}
  \caption{The values of the coefficients of eqn~\ref{3gauss}}
  \begin{tabular}{@{}cc|c|cc@{}}
   \hline
    & Coefficient & \\
    a & b & c \\
   \hline
   58000 & $-0.073$ & $-1.1$\\
   \hline
   90000 & $0.11$ & $-0.71$\\
   \hline
   380 & $3.6$ & $0.82$\\
  \hline
  \label{tab:valcoeff}
  \end{tabular}
 \end{minipage}
\end{table}
Fig.~\ref{galmodel_comparison} compares the $t_{E}$ distribution with catalog simulations using the online form of the Besan\c{c}on Galaxy model by \cite{b54}. Our simple Galactic Model reproduces the timescale distribution slightly better, primarily because the stellar mass function of our model extends to masses in the brown dwarf range. This was achieved at the cost of neglecting the galaxy evolution scenarios of the Besan\c{c}on model capable of reproducing the observed number of microlensing events \citep{b55}. 

Our Galactic Model can be approximated as a multivariate normal distribution for the logarithmic parameters lens mass, Einstein time and Einstein radius $\bmath{p}=\left(\log m_L,\log t_{E},\log R_{E}\right)$:
\begin{equation}
P(\bmath{p}) \propto \exp\left[\frac{1}{2} \left( \bmath{p} -\left<\bmath{p}\right> \right)^T \textbfss{C}^{-1} \left(\bmath{p}-\left<\bmath{p}\right>\right)\right].
\label{multivar_galmodel}
\end{equation}
The parameters of this expression are summarized in Table~\ref{galmodel_pars} and used for all further conclusions. 

\begin{figure}
\includegraphics[width = 0.5\textwidth]{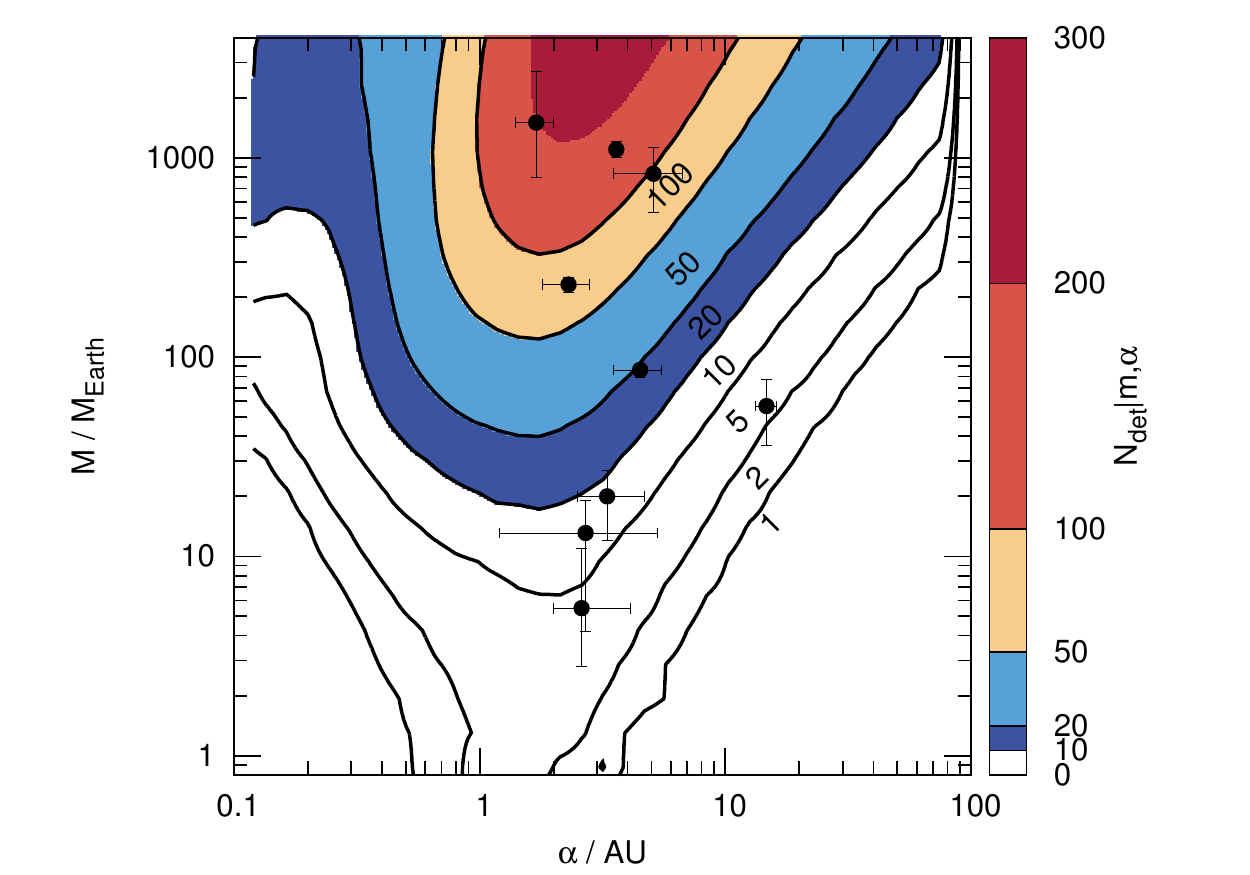}
\caption{Expected number of planet detections (N$_{det}$) simulated for the well-characterized sample of 2433 PSPL events based on their timescale and the Galactic Model in Table~\ref{galmodel_pars}. We assume that each star has a planet of the specified mass at each value of the orbital radius $\alpha$. Filled circles show the locations of previously published microlensing planets. }
\label{combined_prob_AU}
\end{figure}

\begin{table}
\centering
\begin{minipage}{140mm}
\caption{Parameters of the multivariate Gaussian approximation.}
\label{galmodel_pars}
\begin{tabular}{cc}
\hline
         $\left<\log m_L\right>$ & -0.184022\\
          $\left<\log t_E\right>$ &  1.38985\\
          $\left<\log R_E\right>$ &  0.358442\\
          $C_{11}$ & 0.472719\\
          $C_{22}$ & 0.213632\\
          $C_{33}$ & 0.145672\\
          $C_{12}$ & 0.250874\\
          $C_{13}$ & 0.240350\\
          $C_{23}$ & 0.153768\\
         \hline
     \end{tabular}
     \end{minipage}
    \end{table} 

We proceed by convolving our distribution of expected number of detections with Eq.~\ref{multivar_galmodel} and assume that each star has the same chance of hosting a planet and that each planet is uniformly distributed in mass and orbital separation.  

The result of transforming the orbital radii from $R_E$ to AU and mass ratios $q$ to planet masses in $m_{\oplus}$ for the entire set of 2433 events is shown in Fig~\ref{combined_prob_AU}. The published microlensing planets discovered in the OGLE sample in the years 2003 to 2008 are denoted by filled circles\footnote{\citep{b70,b72,b73,b74,b76,b71,b77,b78}}. Fig.~\ref{combined_prob_AU} was generated using only OGLE-III survey data, our Galactic Model and the sensitivities derived from our PSPL fits.

\section{Summary}
\label{sec_sum}
We arrived at an estimate of the planet detection efficiency of the OGLE-III survey from the analysis of an initial sample of 3084 light curves. After we assessed the quality of the data and removed events where the parameters were too loosely constrained, we retained 2433 light curves and used them to estimate the survey sensitivity to planets of different mass ratios at different separations from their host stars. To represent the resulting distribution in more sensible physical units, we employed a Galactic model to convert mass ratios $q$ and projected separations $\alpha/R_E$ to planet masses $m_{\oplus}$ and $\alpha/$AU respectively. The survey sensivitity peaks at ~1-4AU for low mass planets, shifting only slightly to larger separations for higher masses. This result is available as a downloadable fits image at {\it http://robonet.lcogt.net/downloads/planet\_matrix.fits}.

Previously published microlensing planet detections using a combination of survey and follow-up data with discovery dates in the 2003-2008 period feature primarily in the lower sensitivity area of Fig.~\ref{galmodel_comparison}. This suggests that smaller planets are considerably more common than more massive ones. Our results can be used in conjunction with a careful reanalysis of planet candidate events in the OGLE-III survey to place constraints on the abundance of planets orbiting stars several kiloparsec away. A detailed derivation of the planetary mass function will follow in a future work.

{\citet{b79} recently presented a statistical analysis of 224 events observed over four seasons by the OGLE, MOA and Wise microlensing surveys, where they found that 55$^{+34}_{-22}$\% of microlensing events host a planetary companion at or beyond the snow-line. This frequency is compatible with the one estimated by \citet{b49}, provided
the distributions are scaled to the same range of physical units. They also find that Neptune-mass planets are $\sim$10 times more common than Jupiter-mass planets, consistent with what our results indicate.}

\section*{Acknowledgments}
We thank O. Gressell for his advice and support in using the Queen Mary University of London Astronomy Computer Cluster. KH, DB, MD and MH are supported by NPRP grant NPRP-09-476-1-78 from the Qatar National Research Fund (a member of Qatar Foundation). KH acknowledges support from STFC grant ST/M001296/1.
NK acknowledges an ESO Fellowship. The research leading to these results has received funding from the European Community's Seventh Framework Programme (/FP7/2007-2013/) under grant agreement No 229517. This publication was made possible by NPRP grant no.X-019-1-006 from the Qatar National Research Fund (a member of Qatar Foundation).
CS received funding from the European Union Seventh Framework Programme (FP7/2007-2013) under grant agreement no.268421.


\bsp

\label{lastpage}

\newpage

\appendix
\section{Bayesian priors used}
For $A_0, t_E$ we assume a uniform prior in $\log A_0$ and a Gaussian prior in $\log t_E$ respectively.
\begin{equation}
P(A_0) \propto \frac{1}{A_0} exp\left[-\frac{A_0}{\left< A_0 \right>} \right]
\end{equation}

\begin{equation}
P(t_E) \propto \frac{1}{t_E} exp\left[-\frac{1}{2}\left(\frac{\log t_E - \left< \log t_E \right>}{\sigma (\log t_E)}\right)^2\right]
\end{equation}

\end{document}